\documentclass[twocolumn]{aastex631}

\usepackage{aas_macros}
\usepackage{xcolor}
\usepackage{graphicx,hyperref}
\usepackage{dcolumn}
\usepackage{bm}
\usepackage{amsmath}
\usepackage{float}
\usepackage{natbib}

\usepackage{color}

\begin{document}

\title{Neutrino-dominated relativistic viscous accretion flows around rotating black holes with shocks}

\author{Amit Kumar}
\affiliation{Indian Institute of Technology Guwahati, Guwahati 781039, Assam, India}
\correspondingauthor{sbdas@iitg.ac.in}
\email{kamit@iitg.ac.in}

\author{Sayan Chakrabarti}
\affiliation{Indian Institute of Technology Guwahati, Guwahati 781039, Assam, India}
\email{sayan.chakrabarti@iitg.ac.in}

\author[0000-0003-4399-5047]{Santabrata Das}
\affiliation{Indian Institute of Technology Guwahati, Guwahati 781039, Assam, India}
\email{sbdas@iitg.ac.in}

\begin{abstract}

We investigate the relativistic, viscous, advective, neutrino-dominated accretion flows (NDAFs) around rotating stellar mass black holes, incorporating neutrino cooling. By adopting an effective potential to describe the spacetime geometry around the rotating black holes, we self-consistently solve the governing NDAF equations to obtain global transonic accretion solutions. Our findings indicate that, depending on the model parameters, namely energy ($\varepsilon$), angular momentum ($\lambda$), accretion rate ($\dot{m}$), viscosity ($\alpha$) and black hole spin ($a_{\rm k}$), NDAFs may harbor standing shocks where the Rankine-Hugoniot shock conditions (RHCs) are satisfied. Utilizing these shock-induced NDAF solutions, we compute the neutrino luminosity ($L_{\nu}$) and neutrino annihilation luminosity ($L_{\nu \bar{\nu}}$) across a wide range of model parameters. We further calculate maximum neutrino luminosity ($L_{\nu}^{\rm max}$) and neutrino annihilation luminosity ($L_{\nu \bar{\nu}}^{\rm max}$) resulting in $L_{\nu}^{\rm max} \sim 10^{51-53}$ erg s$^{-1}$ ($10^{48-51}$ erg s$^{-1}$) and $L_{\nu \bar{\nu}}^{\rm max} \sim 10^{48-52}$ erg s$^{-1}$ ($10^{42-49}$ erg s$^{-1}$) for $a_{\rm k}=0.99$ (0.0). These findings suggest that shocked NDAF solutions are potentially promising to explain the energy output of gamma-ray bursts (GRBs). We employ our NDAF model formalism to elucidate $L^{\rm obs}_{\nu \bar{\nu}}$ for five GRBs with known redshifts and estimate their accretion rate (${\dot m}$) based on the spin ($a_{\rm k}$) of the central source of GRBs under consideration.
\end{abstract}

\keywords{\href{http://astrothesaurus.org/uat/14}{Accretion (14)}; \href{http://astrothesaurus.org/uat/1963}{Hydrodynamics (1963)}; \href{http://astrothesaurus.org/uat/159}{Black hole physics (159)}; \href{http://astrothesaurus.org/uat/2086}{Shocks (2086)}; \href{http://astrothesaurus.org/uat/629}{Gamma-ray bursts (629)}} 

\section{\label{Introduction}Introduction}

Gamma-ray bursts (GRBs) are among the most energetic events in the universe, exhibiting an isotropic distribution across the sky \cite[]{Meszaros-2001}. They are also the most luminous explosions known, with luminosities reaching up to $10^{\rm 54}\rm{~ erg~ s^{-1}}$. Hence, the origins and underlying physical mechanisms of GRBs are the focus of extensive research.

In general, gamma-ray bursts (GRBs) are classified based on their duration into two categories: short gamma-ray bursts (SGRBs), which last less than 2 seconds, and long gamma-ray bursts (LGRBs), which last more than 2 seconds \cite[]{Kouveliotou-etal-1993}. From the perspective of energy release, accretion onto black holes is considered an efficient mechanism \cite[]{Frank-etal2002}. During accretion, matter revolves around a compact object like a black hole (BH) or a neutron star (NS) and converts its gravitational energy into radiation throughout the process. Indeed, the energy output of GRBs requires accretion rates on the order of a fraction of a solar mass per second up to several solar masses per second. 
  
The probable progenitors of the SGRBs are thought to originate from the mergers of two NSs or a binary NS-BH system \cite[]{Eichler-etal-1989, Paczynski-1991, Narayan-etal-1992}, while LGRBs are mainly associated with the gravitational collapse of massive stars \cite[]{Woosley-1993}. These scenarios often result in a central compact object surrounded by a hyperaccretion disk, which is a plausible candidate for the central engine of GRBs.  In such systems, the extremely high accretion rate causes the inner region of the hyperaccretion disk to become extremely dense ($\rho \sim 10^{\rm 8} - 10^{\rm 12} ~\rm{g~cm^{-3}}$) and hot ($T \sim 10^{10}-10^{11}~\rm{K} $), rendering it optically thick ($\tau_{\gamma}>>1$). This opacity traps photons within the disk, preventing them from contributing to the high-energy emissions associated with GRBs. In contrast, a large number of energetic neutrinos can be emitted from the surface of the disk, which carries away the dissipated energy of the accreted matter. Eventually, the annihilation of neutrinos and anti-neutrinos can produce a relativistic electron-positron pair-dominated outflow, potentially powering a GRB. In the late nineties, \cite{Popham-etal-1999} first proposed the hyperaccreting neutrino-dominated accretion flow (NDAF) model and demonstrate that NDAFs around rotating stellar-mass black holes could serve as a viable central engine for gamma-ray bursts (GRBs). Since then, extensive efforts have been made in the literature to investigate different aspects of NDAF models in explaining GRB characteristics \cite[]{Narayan-etal-2001,DiMatteo-etal-2002,Kohri-Minesgige-2002,Rosswog-etal-2003,Kohri-etal-2005,Kawanaka-Mineshige-2007,Liu-etal-2007,Janiuk-etal2007,Zalamea-Beloborodov2011,Xue-etal-2013,Liu-etal-2017,Chen-etal2022,Wei-Liu2022}. It is important to note that most of the aforementioned studies assume the flow to exhibit Keplerian motion during accretion, and the investigation of transonic behavior including shocks in neutrino-dominated accretion flows remains largely unexplored.

Motivated by this, in this work, we develop a formalism to study the steady, axisymmetric, neutrino-dominated accretion flow (NDAF) around rotating stellar-mass black holes. To account for gravitational effects, we utilize an effective potential that mimics the spacetime geometry surrounding the black hole. We adopt an equation of state that encompasses both gas and radiation pressure. Following \cite[and references therein]{Chakrabarti-Das2004}, we adopt a mixed shear stress prescription to incorporate the effects of viscosity, which governs the transport of angular momentum and heating of matter in NDAFs. Furthermore, the flow is cooled as it accretes toward the black hole, with neutrino cooling being the dominant radiative mechanism. Considering all these, we self-consistently solve the hydrodynamic equations that govern the dynamics of the neutrino-dominated accreting matter around stellar-mass rotating black hole and obtain global transonic solutions for a set of model parameters, such as energy ($\varepsilon$), angular momentum ($\lambda$), accretion rate ($\dot{m}$), viscosity ($\alpha$) and black hole spin ($a_{\rm k}$). We find that during the course of accretion, NDAFs may undergo shock transitions where the Rankine-Hugoniot Conditions (RHCs) \cite[]{Landau-Lifshitz-1959} are satisfied. Such shock-induced global solutions are examined both theoretically and numerically in various accretion environments around weakly ($a_{\rm k}\rightarrow 0$) as well as rapidly ($a_{\rm k}\rightarrow 1$) rotating black holes \cite[]{Fukue-1987,Chakrabarti-1989,Chakrabarti-Molteni1993,Yang-Kafatos1995,Ryu-etal1997,Lu-etal1999,Das-etal2001,Becker-Kazanas-2001,Chakrabarti-Das2004,Das2007,Das-etal-2009,Das-etal2014,Sarkar-Das2016,Okuda-Das2015,Aktar-etal2017,Sukova-etal2017,Dihingia-etal-2018,Dihingia-etal2018a,Sarkar-etal2018,Palit-etal2019,Dihingia-etal2019,Okuda-etal2019,Sen-etal2022,Singh-Das2024,Mitra-Das2024}. We calculate the relevant flow variables and analyze their dependence on the accretion rate ($\dot m$). Furthermore, using the shocked NDAF solutions, we calculate the neutrino luminosity ($L_{\nu}$) and neutrino annihilation luminosity ($L_{\nu \bar{\nu}}$), and examine how they vary with the model parameters. Finally, we discuss the astrophysical relevance of this work in explaining the GRB jet luminosities powered by the annihilation of neutrinos escaping from the disks.

This paper is organized as follows. In Section \ref{Gov_equations}, we outline the model assumptions and describe the governing equations for studying NDAFs. In Section \ref{Results}, we present the global NDAF solutions, both in the presence and absence of shocks. Section \ref{astrophysical_implications} discusses the astrophysical implications of our findings. Finally, in Section $\ref{conclusions}$, we present concluding remarks.

\section{\label{Gov_equations}Governing equations}

We consider steady, axisymmetric, relativistic, viscous, advective, neutrino-dominated accretion flows (NDAFs) around rotating stellar mass black hole. Because of the asymmetry, we use cylindrical coordinate system with black hole located at the origin and choose $G=M_{\rm {BH}}=c=1$, where $G$ is the gravitational constant, $M_{\rm BH}$ is the mass of the black hole and $c$ refers the speed of light. In this system, the radial coordinate, angular momentum, and energy of NDAF are expressed in units of $G M_{\rm{BH}}/c^2$, $G M_{\rm {BH}}/c$, and $c^2$, respectively. 

Based on the above considerations, we formulate the following governing equations for NDAF, which are described below.  
 
\noindent ($1$) Radial momentum equation:
\begin{equation}
    v\frac{dv}{dx}+\frac{1}{\rho}\frac{dP}{ dx}+\frac{d \Phi^{\rm {eff}}}{dx}=0,
    \label{rad_mom_eq}
\end{equation}
where $x$, $v$ and $\rho$ are radial distance, radial velocity and mass density of flow, respectively. The total pressure $P$ of the flow is the sum of the gas pressure ($P_{\rm gas}$) and radiation pressure ($P_{\rm rad}$). The gas pressure is expressed as $P_{\rm gas}=\rho k_{\rm B} T /m_{\rm p}$, where $k_{\rm{B}}$, $T$ and $m_{\rm p}$ represent the Boltzmann constant, temperature of the flow and mass of the ion. The radiation pressure is given by $P_{\rm rad}=11 \bar{a} T^4/12$. Here, $\bar{a}$ is radiation constant and the factor $11/12$ accounts the contribution of relativistic electron-positron pairs \cite[]{DiMatteo-etal-2002}. In equation (\ref{rad_mom_eq}), $\Phi^{\rm{eff}}$ refers effective potential that mimics the spacetime geometry around a rotating black hole \cite[]{Dihingia-etal-2018} and is given by,
\begin{equation}
    \Phi^{\rm eff}=\frac{1}{2} \ln\left[\frac{x \Delta}{a^2_{\rm{k}}(x+2)-4 a_{\rm k} \lambda+x^3-\lambda^2(x-2)}\right],\label{potential}
\end{equation}
where $\lambda$ denotes the specific angular momentum of the flow, $a_{\rm{k}}$ is the spin of the black hole and $\Delta = x^2-2x+a^2_{\rm{k}}$.

\noindent ($2$) Mass conservation equation:
\begin{equation}
    \dot M=2\pi v\Sigma\sqrt{\Delta},\label{mass_cons_eq}
\end{equation}
where $\dot M$ is the accretion rate which remains constant throughout the flow and $\Sigma~(= 2\rho H)$ represents the vertically integrated mass density of flow \cite[]{Matsumoto-etal-1984}. The quantity $H$ denotes the half thickness of the disk and is defined as \cite[]{Riffert-Herold-1995,Peitz-Appl1997},
$$
H=\sqrt{\frac{Px^3}{\rho \mathcal{F}} }; \quad \mathcal{F}=\frac{1}{(1-\Omega\lambda)}\frac{(x^2+a^2_{\rm{k}})^2+2 \Delta a^2_{\rm{k}}}{(x^2+a^2_{\rm{k}})^2-2 \Delta a^2_{\rm{k}}},
$$
where $\Omega~\left[=(2 a_{\rm k}+\lambda(x-2))/(a_{\rm k}^2(x+2)-2 a_{\rm k}\lambda +x^3)\right]$ is the angular velocity of flow. In this work, we write $\dot{M}$ in unit of solar mass per second and is expressed as $\dot{m}=\dot{M}/(M_\odot~{\rm s}^{-1})$.

\noindent ($3$) Azimuthal momentum equation:
\begin{equation}
    \Sigma v x \frac{d\lambda}{dx}+\frac{d}{dx}(x^2 W_{{x\phi}})=0,\label{azimuthal_eq}
\end{equation}
where $W_{x\phi}~[= -\alpha (W+\Sigma v^2)]$ is the $x \phi$ component of the viscous stress \cite[]{Chakrabarti-Molteni-1995}, $W~(=2 P H)$ is the vertically integrated pressure and $\alpha$ is the viscosity parameter. 
  
\noindent ($4$) Entropy generation equation:
\begin{equation}
    \Sigma v T\frac{ds}{dx}=\Sigma v  \left(\frac{du}{dx}-\frac{P}{\rho^2}\frac{d\rho}{dx}\right)=Q^{-}-Q^{+},\label{entrop_gen_eq}
\end{equation}
where $s$ represent the specific entropy of the flow. The specific internal energy ($u$) of the flow is expressed as 
\begin{align}
    u=\frac{P_{\rm{gas}}}{\rho(\gamma-1)}+\frac{3P_{\rm{rad}}}{\rho},\label{total_internal_energy}
\end{align}
where $\gamma$ denotes the adiabatic index which we keep fixed as $4/3$ throughout this study. In equation (\ref{entrop_gen_eq}), $Q^{+}$ and $Q^{-}$ represent the energy gain due to viscous heating and energy loss due to neutrino cooling, respectively. Following \cite[]{Chakrabarti-Das2004}, the heating term $Q^{+}$ is expressed as
\begin{align}
    Q^{+}=Q_{\rm vis}=-\alpha(\Sigma v^2+W)x\frac{d \Omega}{dx},\label{viscous_heating}
\end{align}
whereas, the cooling due to loss of neutrinos is obtained as \cite{DiMatteo-etal-2002},
\begin{equation}
     Q^{-}= Q_\nu=\sum_{i}\frac{(7/8)\sigma T^4}{(3/4)[\tau_{\nu_i}/2+1/\sqrt{3}+1/(3\tau_{a,\nu_i})]},\label{neutrino_cooling}
\end{equation}
where $\tau_{\nu_i}~[=\tau_{a,\nu_i}+\tau_{s,\nu_i}]$ is the sum of absorptive and scattering optical depths corresponding to each neutrino flavour. The index $i$ refers to both electron type neutrinos $\nu_{\rm e},\bar{\nu}_{\rm e}$ and heavy lepton neutrinos $\nu_{\mu},\bar{\nu}_{\mu}, \nu_{\tau}$ and $\bar{\nu}_{\tau}$, while the subscript $s$ and $a$ are used to denote scattering and absorption, respectively. It is worth mentioning that equation (\ref{neutrino_cooling}) is valid in both optically thin and thick limit for neutrinos. Neutrinos can be produced through neutronization and thermal emission via various types of reactions, which are briefly described below \cite{DiMatteo-etal-2002}.

\noindent (I) Electron-positron pair annihilation  ($e^{-}+e^{+}\rightarrow \nu_{i}+\bar{\nu}_{i}$) \cite[]{Liu-etal-2007}, where the neutrino cooling rate per unit volume is given by,
\begin{align}
     q^-_{\rm{e^{-}e^{+}\rightarrow\nu_{e}+\bar{\nu}_{e}}}&\approx 3.4\times 10^{33}T^9_{11}\,\rm{ erg \, cm^{-3}\, s^{-1}}\label{q_epem>nue+nueb}, \\
     q^-_{\rm{e^{-}e^{+}\rightarrow\nu_{\mu}+\bar{\nu}_{\mu}}}&= q_{\rm{e^{-}e^{+}\rightarrow\nu_{\tau}+\bar{\nu}_{\tau}}}\notag \\
     &\approx 0.7\times10^{33}T^9_{11}\,\rm{ erg \, cm^{-3}\, s^{-1}},\label{q_epem>numu+numub}
\end{align}
where the temperature of the flow is expressed as $T=T_{11} \times 10^{11}{\rm K}$, $T_{11}$ being scaled temperature.

\noindent (II) The electron-positron pair capture on nuclei occurs through the reactions $(p+e^{-}\rightarrow n+\nu_{\rm e}$ and $n+e^{+}\rightarrow p+\bar{\nu}_{\rm e}$), a process known as neutronization \cite[]{DiMatteo-etal-2002}. The corresponding cooling rate per unit volume is given by,
\begin{equation}
q^{-}_{\rm{eN}}=q^{-}_{\rm{e^{-}p}}+q^{-}_{\rm{e^{+}n}}=9.0\times10^{33}\rho_{10}T^6_{11}\,\rm{erg \, cm^{-3} \,s^{-1}},\label{q_electronpositroncapture}
\end{equation}
where $\rho =\rho_{10} \times 10^{10}~ {\rm g}~ {\rm cm}^{-3}$, $\rho_{10}$ being scaled density.

\noindent (III) The nucleon-nucleon bremsstrahlung process, described by the reaction $n+n\rightarrow n+n+\nu_{i}+\bar{\nu}_{i}$, contributes as follows \cite{Hannestad-Raffelt-1998}:
\begin{equation}
    q^-_{\rm{brem}}\approx1.5\times10^{27}\rho^2_{10} T^{5.5}_{11}\, \rm{erg \, cm^{-3} \,s^{-1}}\label{q_brem}.
\end{equation}

\noindent (IV) The plasmon decay is associated with the decay rate of transverse plasmons, indicating that standard photons interact with the electron gas via the decay process $\tilde\gamma\longrightarrow \nu_{\rm{e}}+\bar\nu_{\rm{e}}$. The expression for $q^-_{\rm{plasmon}}$ is given by \cite{Ruffert-etal-1996},
\begin{align}
 q^-_{\rm{plasmon}} \approx & 1.5\times 10^{32} ~ T^9_{11}\gamma^6_{\rm{p}}\exp(-\gamma_{\rm p}) \nonumber \\
 & \times \left(2+ 2 \gamma_{\rm p} + \gamma^2_{\rm p} \right)
 \, \rm{erg \, cm^{-3} \,s^{-1}}\label{q_palsmon}  
\end{align}
where $\gamma_{\rm p}$ is defined as $\gamma_{\rm p}=5.565\times10^{-2}[(\pi^2+\eta^2_{\rm e})/3]^{1/2}$ 
and $\eta_{\rm e}=\mu_{\rm e}/k_{\rm B} T$, $\mu_{\rm{e}}$ being the electron chemical potential.

Now, we introduce the various absorption optical depths for different neutrino species defined as, 
\begin{equation}
    \tau_{a, \nu_{\rm e}} =\frac{(q^-_{\rm{e^{-}e^{+}}\rightarrow \nu_{e}+\bar{\nu}_{e}}+q^{-}_{\rm{eN}}+q^-_{\rm{brem}}+q^-_{\rm{plasmon}})H}{4(7/8)\sigma T^4},\label{tau_a_e} 
\end{equation} 
\begin{equation}
    \tau_{a, \nu_{\mu}} =\tau_{a, \nu_{\tau}} =\frac{(q^-_{\rm{e^{-}e^{+}}\rightarrow\nu_{\mu}+\bar{\nu}_{\mu}}+ q^-_{\rm{brem}})H}{4(7/8)\sigma T^4}\label{tau_a_mu=tau_a_tau}
\end{equation}

On the contrary, any type of scattering delays the free escape of neutrinos from the disk. Additionally, neutrinos are also scattered by nucleons as well, and the corresponding scattering optical depth is given by \cite{DiMatteo-etal-2002}, 
\begin{equation}
     \tau_{s,\nu_i}=2.7\times10^{-7}T^2_{11}\rho_{10} H.\label{scattering_optical_depth}
\end{equation}

Putting the above absorption and scattering optical depth expressions in (\ref{neutrino_cooling}) and using equations (\ref{rad_mom_eq}), (\ref{mass_cons_eq}), (\ref{azimuthal_eq}),(\ref{entrop_gen_eq}), we obtain the gradient of radial velocity as, 
\begin{equation}
    \frac{dv}{dx}=\frac{\mathcal{N}(x,v,\lambda,\Theta)}{\mathcal{D}(x,v,\lambda,\Theta)}\label{dvdr},
\end{equation} 
where $\Theta=k_{\rm{B}}T/(m_{\rm{e}}c^2)$. Furthermore, the gradient of  angular momentum ($\lambda$) and temperature ($\Theta$) is obtained as,
\begin{equation}
    \frac{d\lambda}{dx}=\lambda_{11}\frac{dv}{dx}+\lambda_{12},\label{dldr}
\end{equation}
and 
\begin{equation}
    \frac{d\Theta}{dx}=\Theta_{11}\frac{dv}{dx}+\Theta_{12},\label{dthetadr}
\end{equation}
where the explicit expressions of $\lambda_{11}, \lambda_{12}, \Theta_{11}$, and $\Theta_{12}$ are given in Appendix A.

The accreting matter typically begins its journey from the outer edge of disk with a very low (subsonic) radial velocity and ultimately crosses the black hole horizon with velocity comparable to the speed of light. Along the streamline, flow must pass through a critical point ($x_{\rm c}$) where it smoothly transits from subsonic to supersonic speeds. At this critical point, the equation ($\ref{dvdr}$) becomes indeterminate in the form $(dv/dx)_{x_{\rm c}}=0/0$, allowing us to derive the critical point condition as \cite[]{Chakrabarti-Das2004}, 
\begin{equation}
    \mathcal{N}(x,v,\lambda,\Theta)_{x_{\rm c}}= \mathcal{D}(x,v,\lambda,\Theta)_{x_{\rm c}}=0. \label{N=D=0}
\end{equation}

Since the flow remain smooth and continuous at each radii, $(dv/dx)$ must be finite all throughout and hence, we apply l$'$H\^{o}pital's rule to evaluate $(dv/dx)_{x_{\rm c}}$. Generally, a physically acceptable critical point that astrophysical flow passes through yields two distinct values: one for accretion $(dv/dx)_{x_{\rm c}}<0$ and another for wind $(dv/dx)_{x_{\rm c}}>0$. The critical point formed closest to the black hole horizon ($x_{\rm h}$) is referred to as the inner critical point ($x_{\rm in}$), while the critical point located farther away is called the outer critical point ($x_{\rm out}$).

\section{Results}
\label{Results}

We self-consistently solve equations (\ref{dvdr}, \ref{dldr}, \ref{dthetadr}) using 4th order Runge-Kutta method to derive the global accretion solutions around the black hole. For this, we specify the global flow parameters $\alpha$, $\dot{m}$ and $a_{\rm{k}}$. Further, we make use of critical point $x_{\rm c}$ and angular momentum ($\lambda_{\rm c}$) at $x_{\rm c}$ as local parameters. Using these parameters, we solve equation (\ref{N=D=0}) to calculate $v_{\rm c}$ and $\Theta_{\rm c}$ at $x_{\rm c}$. Subsequently, we compute the local flow energy ($\varepsilon_{\rm c}$) at $x_{\rm c}$ following $\varepsilon(x)=v^2/2 + h +\Phi^{\rm eff}$, $h$ being the enthalpy of the flow and is given by $h= u+P/\rho$. With these initial values of the flow variables ($i.e$, $x_{\rm c}, \lambda_{\rm c}, v_{\rm c}, \Theta_{\rm c}, \alpha, \dot{m}$), we integrate equations ($\ref{dvdr}$)-($\ref{dthetadr}$) inward from $x_{\rm c}$ to the black hole horizon ($x_{\rm h}$) and outward from $x_{\rm c}$ to the outer edge $x_{\rm edge}=1000$ for a chosen $a_{\rm k}$ values. Finally, we combine these two segments to obtain the global transonic accretion solution, which connects the BH horizon to the outer edge of the disk.

\subsection{Global solutions of NDAF}

\begin{figure}[ht!]
    \centering
       \includegraphics[width=\columnwidth]{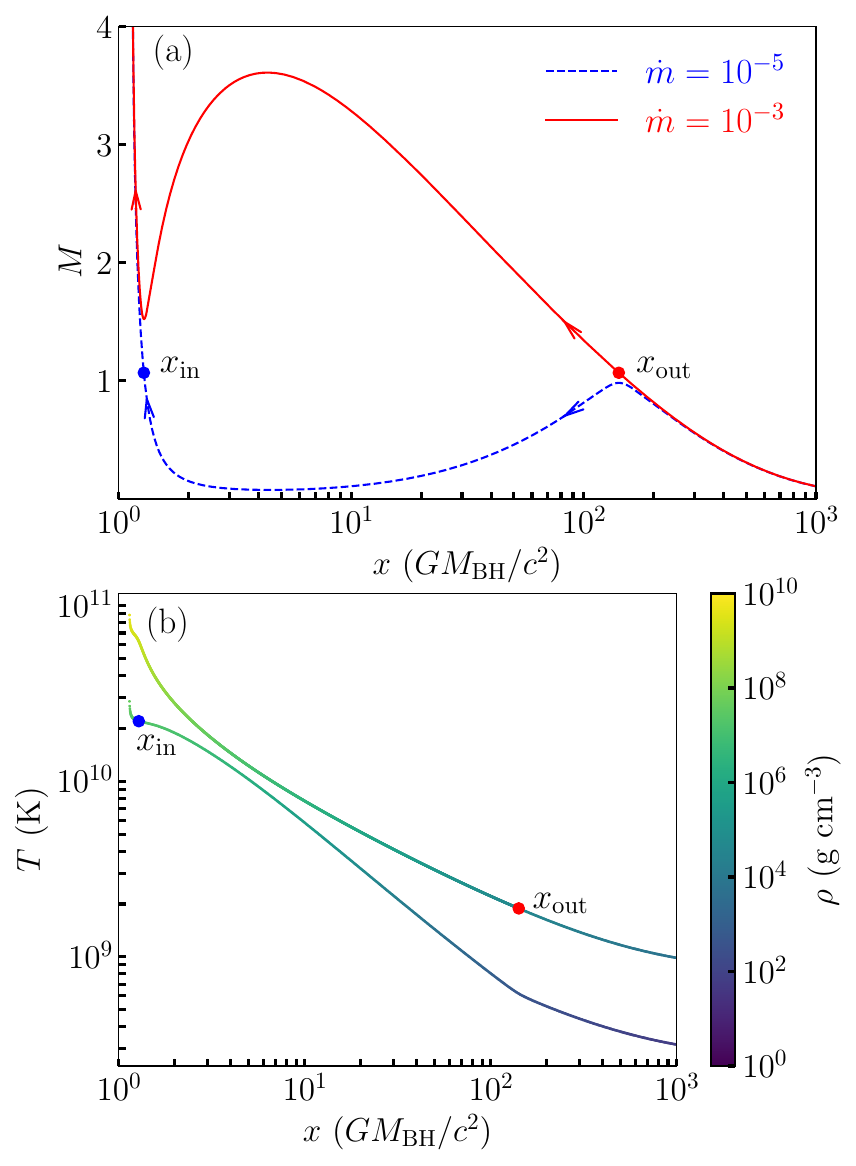}
    \caption{(a) Variation of Mach number ($M=v/C_{\rm s}$) with radial distance ($x$) for different accretion rate  ($\dot{m}$). NDAF in injected from outer edge $x_{\rm{edge}}=1000$ with $\varepsilon_{\rm edge}=4.0668\times 10^{-3}$, $\lambda_{\rm{edge}}=2.4292$, $\alpha=0.001$ around a black hole of spin $a_{\rm{k}}=0.99$. Dashed (blue) and solid (red) curves represent the results for ${\dot m}=10^{-5}$ and $10^{-3}$, respectively. Critical points ($x_{\rm in}$ and $x_{\rm out}$) are marked, and arrows represent the direction of flow motion towards the black hole. (b) Temperature ($T$) variation of NDAF as function of radial distance ($x$) for solutions presented in panel (a). Color indicates the density ($\rho$) variation with $x$ where color bar shows the range of density. See the text for details.}
    \label{fig:fig01}
\end{figure}

In Fig. \ref{fig:fig01}a, we present the examples of accretion solutions where Mach number ($M=v/C_{\rm s}$, $C_{\rm s}$ being sound speed) is plotted with radial distance ($x$). Here, we choose the local flow parameters at inner critical point as $x_{\rm in}=1.2833$ and $\lambda_{\rm in}=2.1022$, and global parameters as $\alpha=0.001$, and $a_{\rm k}=0.99$. For $\dot{m}=10^{-5}$, we find $v_{\rm in}=0.3194$ and $\Theta_{\rm in}=3.7069$ at $x_{\rm in}$ yielding $\varepsilon_{\rm in}=3.5054 \times 10^{-3}$, and the obtained result is plotted using dashed (blue) curve. We note the flow variables at the outer edge of the disk at $x_{\rm edge}=1000$ as $v_{\rm edge}=3.8047 \times 10^{-3}$, $\lambda_{\rm edge}=2.4292$ and $\Theta_{\rm edge}=0.0533$, and compute the local energy $\varepsilon_{\rm edge}= 4.0668 \times 10^{-3}$. It is worth mentioning that one can get the identical accretion solution by integrating equations ($\ref{dvdr}$)-($\ref{dthetadr}$) from $x_{\rm edge}$ towards the horizon using these outer boundary values of the accretion flow ($x_{\rm edge},\varepsilon_{\rm edge},\lambda_{\rm edge}$). Next, we increase the accretion rate as $\dot{m}=10^{-3}$ while keeping the remaining flow variables ($\varepsilon_{\rm edge}$, $\lambda_{\rm edge}$, $\alpha$, and $a_{\rm k}$) fixed at $x_{\rm edge}=1000$ and compute the global transonic accretion solution by suitably tuning the velocity and temperature as $v_{\rm edge}=3.8269 \times 10^{-3}$ and $\Theta_{\rm edge}=0.0356$ that satisfy the critical point conditions (equation (20)). We observe that accretion solution alters its character as it passes through the outer critical point ($x_{\rm out}=141.7357$) instead of inner critical point ($x_{\rm in}$) with $\lambda_{\rm out} = 2.1092$, $v_{\rm out}= 0.0526$ and $\Theta_{\rm out}= 0.3180$. Thereafter, we again integrate equations ($\ref{dvdr}$)-($\ref{dthetadr}$) inward from $x_{\rm out}$ to the horizon ($x_{\rm h}$) and obtain the global transonic accretion solution in the range $x_{\rm h} < x \le x_{\rm edge}$. This result is plotted using solid (red) curve. In the figure, arrows indicate the direction of flow motion towards the black hole and critical points ($x_{\rm in}$ and $x_{\rm out}$) are marked. In Fig. \ref{fig:fig01}b, we show the temperature profiles of the NDAF solutions illustrated in Fig. \ref{fig:fig01}a. For ${\dot m} = 10^{-5}$, the flow temperature reaches approximately $T \sim 3 \times 10^{10}$ K near the horizon, while for ${\dot m} = 10^{-3}$, the temperature exceeds $T \sim 9 \times 10^{10}$. The corresponding density profiles ($\rho$) are represented by color, with the colorbar on the right indicating the range of flow density. It is evident from the figure that the flow density remains very high in the inner part of the disk for both cases, with values around $\rho \sim 10^{8-10}$ g cm$^{-3}$.

\subsection{Shock-induced NDAF solutions}

\begin{figure}
    \centering
       \includegraphics[width=\columnwidth]{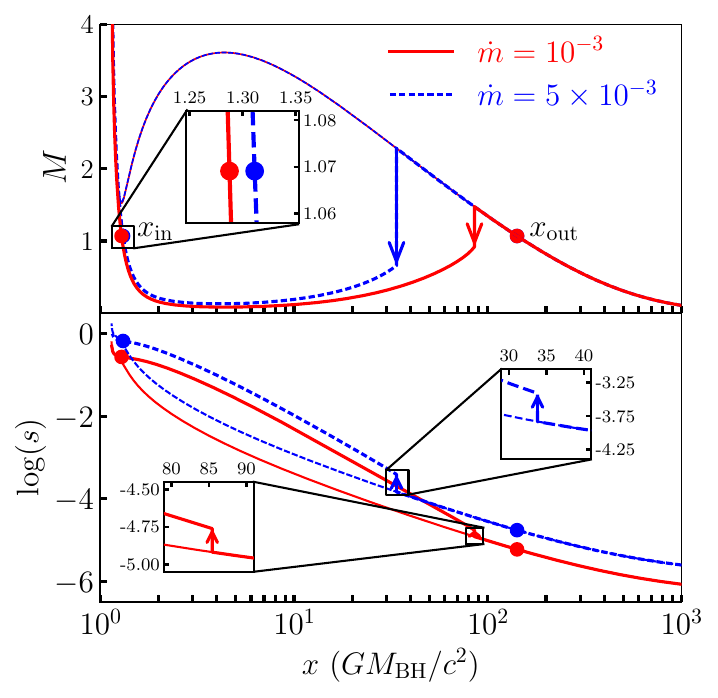}
    \caption{Examples of shock-induced global NDAF solutions obtained for different accretion rate ($\dot m$). Solid (red) and dashed (blue) curves denote results obtained for ${\dot m}=10^{-3}$ and $5\times 10^{-3}$, where vertical arrows indicate the shock radii $x_{\rm s}=85.4376$ and $x_{\rm s}=33.8093$. Here, the other model parameters are chosen to be the same as those used in Fig. \ref{fig:fig01}. See the text for details.}
    \label{fig:fig02}
\end{figure}

In this section, we present for the first time the shock-induced global accretion solution in NDAFs. As indicated in Fig. \ref{fig:fig01}, the flow begin with negligible radial velocity, gradually increasing as it moves toward the black hole. During this journey, the flow undergoes a sonic state transition from subsonic to supersonic as it crosses the outer critical point ($x_{\rm out}$) before entering in to the black hole. Interestingly, due to the rotation of matter, the flow experiences centrifugal repulsion during the accretion process. This leads to an accumulation of infalling matter, forming an effective boundary layer around the black hole. It is important to note that this accumulation of matter does not continue indefinitely as the centrifugal barrier triggers the discontinuous shock transition in the flow variables once limiting threshold is reached \cite[]{Fukue-1987, Chakrabarti-1989, Das-Chakrabarti-2004}. Indeed, the global accretion solutions containing shock waves are thermodynamically preferred compared to shock-free solutions due to its higher entropy content \cite{Becker-Kazanas-2001}.

When accretion flow experiences shock transition, Rankine-Hugoniot conditions \cite[RHCs,][]{Landau-Lifshitz-1959} must be satisfied. In the context of vertically integrated accretion flows, the RHCs are expressed as follows: (a) continuity of mass flux $\dot{M}_{+}= \dot{M}_{-}$, (b) continuity of energy flux $\varepsilon_{+}=\varepsilon_{-}$, and (c) continuity of momentum flux $\Pi_{+}= \Pi_{-}$. Here, $\Pi~(=W+\Sigma v^2)$, denotess the vertically integrated total pressure and the quantities with subscripts `$-$' and `$+$' represent the flow variable just before and after the shock transition.

To examine the shock transition\footnote{See Appendix-C for details}, we make use of the accretion solutions passing through the outer critical point ($x_{\rm out}$) as shown in Fig. \ref{fig:fig01}. For this solution, we observe that after crossing the outer critical point, the supersonic flow undergoes discontinuous shock transition to the subsonic branch at $x_{\rm s}= 85.4376$, where RHCs are satisfied. After the shock, the slowly moving flow gradually gains its radial velocity as it proceeds inward and ultimately enters into the black hole after smoothly crossing the inner critical point at $x_{\rm in}=1.2969$. The result of shock-induced global accretion solution is depicted in the upper panel of Fig. \ref{fig:fig02} using solid (red) curve, where vertical arrow indicates the shock transition radius. To examine the role of accretion rate ($\dot{m}$) on shock formation, we choose $\dot{m}=5\times 10^{-3}$, keeping other model parameters unchanged ($i.e.$, $\varepsilon_{\rm edge}=4.0668\times 10^{-3}$, $\lambda_{\rm{edge}}=2.4292$, $\alpha=0.001$, and $a_{\rm k}=0.99$). We find that shock forms at a lower radius $x_{\rm s}=33.8093$, as illustrated by the dashed (blue) curve. This happens because a higher accretion rate increases both the density ($\rho$) (see equation \ref{mass_cons_eq})) and the temperature ($T$) of the flow. Specifically, the increase in $\rho$ and $T$ leads to enhanced neutrino cooling (see equations (\ref{neutrino_cooling})), which pushes the shock front inward, where RHCs are satisfied. In the figure, arrows indicate the overall direction of the flow motion towards the black hole. In the lower panel of Fig. \ref{fig:fig02}, we present the variation of the specific entropy for both shocked and shock-free solutions, shown in the upper panel. Following \cite[]{Chandrasekhar-1939, Das-etal-2009, Mitra-Das2024}, we compute the specific entropy of the flow as $s~\propto P/\rho^{\gamma-1}$. We observe that the post-shock flow possesses higher specific entropy as it increases across the shock front. With this, we emphasize that shocked accretion solutions are thermodynamically preferred over shock-free solutions.

\begin{figure}
    \centering
       \includegraphics[width=\columnwidth]{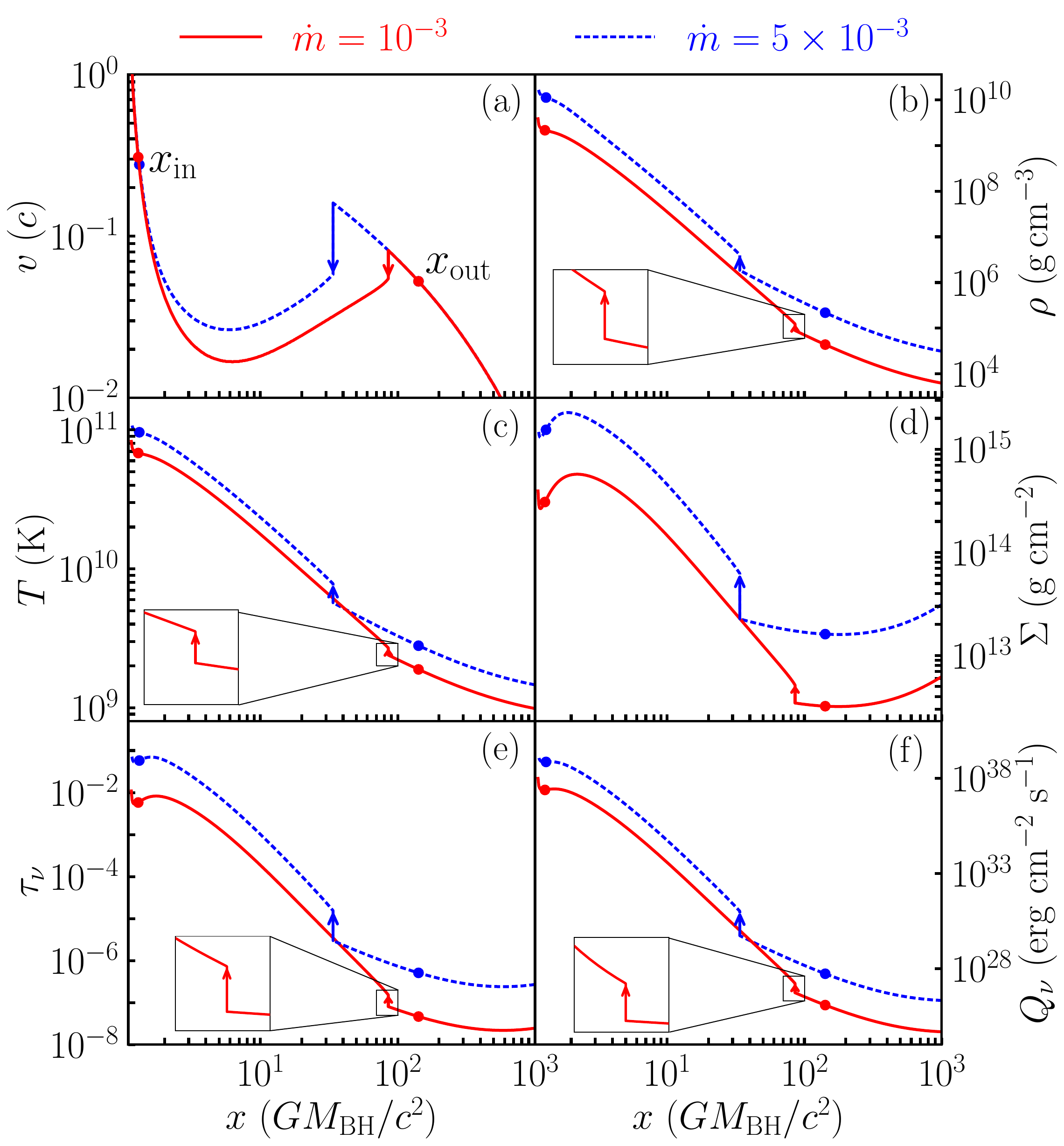}
    \caption{Radial variation of (a) radial velocity $v$, (b) mass density $\rho$, (c) temperature $T$, (d) surface density $\Sigma$, (e) neutrino optical depth and (f) neutrino cooling $Q_{\nu}$ for shocked solutions presented in Fig. \ref{fig:fig02}. Solid (red) and dashed (blue) curves are for ${\dot m}=10^{-3}$ and $5 \times 10^{-3}$, respectively. In each panel, filled circles indicate the critical points and shock transition is shown by vertical arrow. See the text for details.
    }
    \label{fig:fig03}
\end{figure}

In Fig. \ref{fig:fig03}, we depict the radial profiles of the flow variables corresponding to the shock-induced global accretion solutions presented in Fig. \ref {fig:fig02}. Solid (red) and dashed (blue) curves represent flow variables for $\dot{m}=10^{-3}$ and $5 \times 10^{-3}$, respectively. In panel (a), we show the variation of radial velocity ($v$) with radial distance ($x$). We observe that $v$ increases as matter accretes towards the black hole from the outer edge of the disk ($x_{\rm edge}=1000$) and sharply decreases across the shock front as shown by vertical arrows. However, due to extreme gravitational attraction of the black hole, the flow subsequently gains it velocity and ultimately falls into the black hole horizon at supersonic speeds. Panel (b) illustrates the mass density profile $\rho$ as a function of $x$. Due to shock, convergent flow experiences density compression at the post-shock region (referred to as the post-shock corona, or PSC), resulting in an increase in $\rho$. This increase of $\rho$ occurs because the radial velocity of the flow decreases across the shock front, that causes the increase in density in order to conserve the mass flux, which is one of the fundamental condition for shock transition. In panel (c), we display the variation of temperature ($T$) with $x$. A notable jump in temperature occurs across the shock, as the kinetic energy of the pre-shock flow is converted into thermal energy, leading to an increase in the post-shock region. Panel (d) presents the radial variation of surface density ($\Sigma$), revealing that $\Sigma$ rises at PSC due to the density compression occurring at the shock front. In panel (e), we present the neutrino optical depth. It is clear that the flow remains optically thin across the entire length scale, with $\tau_{\nu} < 1$, allowing neutrinos to escape easily from the disk. Finally, panel (f) illustrates the profile of neutrino cooling (in $\rm{erg~cm^{-2}~s^{-1}}$) as a function of radial distance, incorporating contributions from various processes discussed in section \ref{Gov_equations}. Given the higher temperature and density in the post-shock region compared to the pre-shock region, it is evident that the net energy loss will be greater in the post-shock flow.
 
\begin{figure}
    \centering
       \includegraphics[width=\columnwidth]{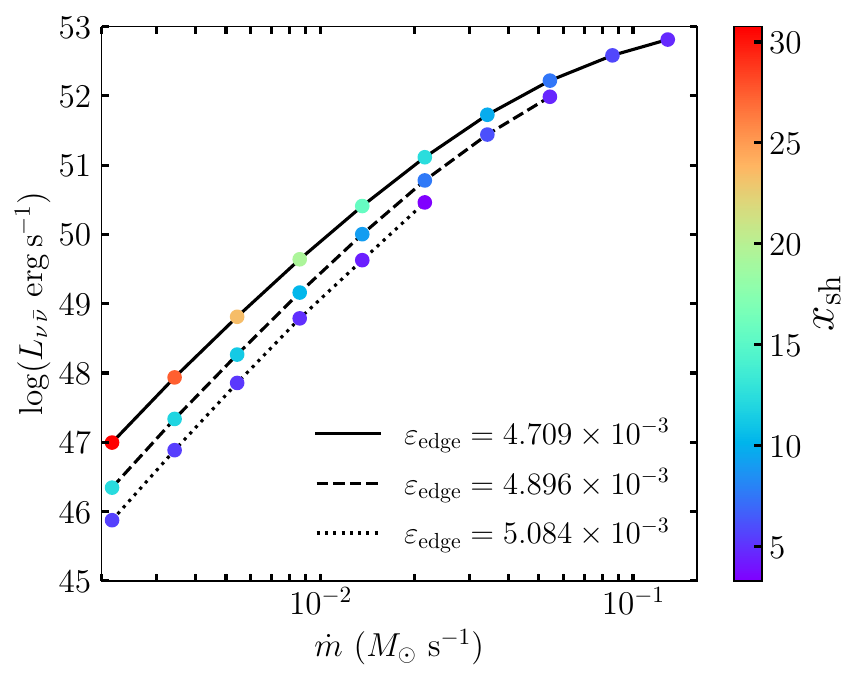}
    \caption{Variation of neutrino annihilation luminosity ($L_{\nu\bar{\nu}}$) with accretion rate ($\dot{m}$) for shocked solutions. Here, we fix the model parameters as $(a_{\rm k }$, $\alpha$, $\lambda_{\rm edge}$) $=$ $(0.99$, $0.001$, $2.5331)$. Filled circles connected by solid, dashed and dotted lines represent results corresponding to $\varepsilon_{\rm edge}=4.709\times 10^{-3}$, $4.834\times 10^{-3}$, $4.959\times 10^{-3}$ and $5.084\times 10^{-3}$, respectively. Colors denote the shock radii. See the text for details.}
    \label{fig:fig04}
\end{figure}

\subsection{Neutrino luminosity ($L_{\rm{\nu}}$) and neutrino annihilation luminosity ($L_{\nu\, \bar{\nu}}$)}

Having established the global accretion solutions for NDAFs, we now turn our attention to estimating the neutrino radiation luminosity prior to the annihilation process. We calculate the neutrino radiation luminosity of the NDAF as \cite{Liu-etal-2007},
\begin{equation}
     L_{\rm{\nu}}=4\pi \int^{x_{f}}_{x_i}Q_{\rm{\nu}}\,x\, dx ,\label{neutrino_luminosity}
 \end{equation}
where $x_i$ denotes the radius just outside the horizon ($x_{\rm h}$), $x_f$ refers the outer edge of the disk ($x_{\rm edge}$) and $Q_{\rm{\nu}}$ is the neutrino cooling rate expressed in unit of erg cm$^{-2}$ s$^{-1}$. Following \cite[]{Ruffert-etal-1997,Popham-etal-1999,Rosswog-etal-2003}, we estimate the neutrino annihilation luminosity by dividing the disk into a grid of cells in the equatorial plane, with each cell characterized by its mean neutrino energy. The $k$-th cell has a mean neutrino energy denoted as $\epsilon^{\rm k}_{\nu_i}$ and a neutrino radiation luminosity of $l_{\nu_i}^{\rm k}$. The distance to a spatial point above (or below) the disk is represented by $d_{\rm k}$. The quantity $l_{\nu_i}^{\rm k}$ is calculated for each neutrino flavor in a cell through the surface integral of neutrino cooling (see equation \ref{neutrino_luminosity}). The angle at which neutrinos from cell $k$ interact with antineutrinos from another cell $k^{\prime}$ is denoted by $\theta_{\rm k,k^{\prime}}$. The energy deposition rate per unit volume (in unit of $\rm{erg~cm^{-3}~s^{-1}}$) at that point is given by,
\begin{align}
    l_{\nu\bar{\nu}}= & \sum_{i} A_{1,i} \sum_{\rm k}\frac{l_{\nu_i}^{\rm k}}{d^{\rm{2}}_{\rm k}} \sum_{\rm k^{\prime}}\frac{l_{\bar{\nu_i}}^{\rm k^{\prime}}}{d^{\rm{2}}_{\rm k^{\prime}}}(\epsilon^{\rm k}_{\nu_i}  +  \epsilon^{\rm k^{\prime}}_{\bar{\nu_i}}   )(1-\cos\, \theta_{\rm k\,k^{\prime}})^2 \nonumber \\
    &+ \sum_{i} A_{2,i} \sum_{\rm k}\frac{l_{\nu_i}^{\rm k}}{d^2_{\rm k}} \sum_{\rm k^{\prime}} \frac{l_{\bar \nu_i}^{\rm k^{\prime}}}{d^2_{\rm k^{\prime}}}\frac{   \epsilon^{\rm k}_{\nu_i}  +  \epsilon^{\rm k^{\prime}}_{\bar{\nu_i}}}{ \epsilon^{\rm k}_{\nu_i}    \epsilon^{\rm k^{\prime}}_{\bar{\nu_i}}}(1-\cos\, \theta_{\rm k\,k^{\prime}}).\label{lnunubar}
\end{align}
In equation (\ref{lnunubar}), the explicit expression of $A_{1,i}$ and $A_{2,i}$ are given in Appendix-B. We calculate the total neutrino annihilation luminosity by integrating over the entire region outside the black hole and the disk, as described in \cite[]{Xue-etal-2013} and is expressed as,
\begin{equation}
L_{\nu \bar{\nu}}= 4\pi \int^\infty_{x_i} \int^\infty_{H} l_{\nu\bar{\nu}} x dx dz,
\label{neutrino_anni_lum}
\end{equation}
where $z$ refers the vertical axis.

In Fig. \ref{fig:fig04}, we present the variation of $L_{\nu\bar{\nu}}$ as a function of accretion rate ($\dot{m}$) for different $\varepsilon_{\rm edge}$. Here, we choose $\lambda_{\rm edge} = 2.5331$, $\alpha = 0.001$ and $a_{\rm k }=0.99$. Filled circles connected with solid, dashed and dotted lines are obtained for $\varepsilon_{\rm edge}=4.709 \times 10^{-3}$, $4.896 \times 10^{-3}$ and $5.084\times 10^{-3}$, respectively. Colors represent the shock radii, with the range indicated by the color bar on the right side of the figure. Notably, when $\varepsilon_{\rm edge}$ is relatively smaller, shock forms over a broader, whereas larger values of $\varepsilon_{\rm edge}$ result in a more limited range for shock formation. We observe that for a given $\varepsilon_{\rm edge}$, $L_{\nu\bar{\nu}}$ increases with $\dot m$. This behavior can be attributed to the fact that the density of the flow increases with the accretion rate. As density rises, neutrino cooling in the flow is enhanced, which in turn leads to an increase in luminosity. On contrary, for a fixed $\dot m$, $L_{\nu\bar{\nu}}$ decreases with the increase of $\varepsilon_{\rm edge}$. This occurs because an increase in $\varepsilon_{\rm edge}$ causes the shock front to shift toward the horizon, resulting in a reduction in the size of the post-shock region (PSC). Consequently, this leads to a decrease in $L_{\nu\bar{\nu}}$.

\begin{figure}
    \centering
       \includegraphics[width=\columnwidth]{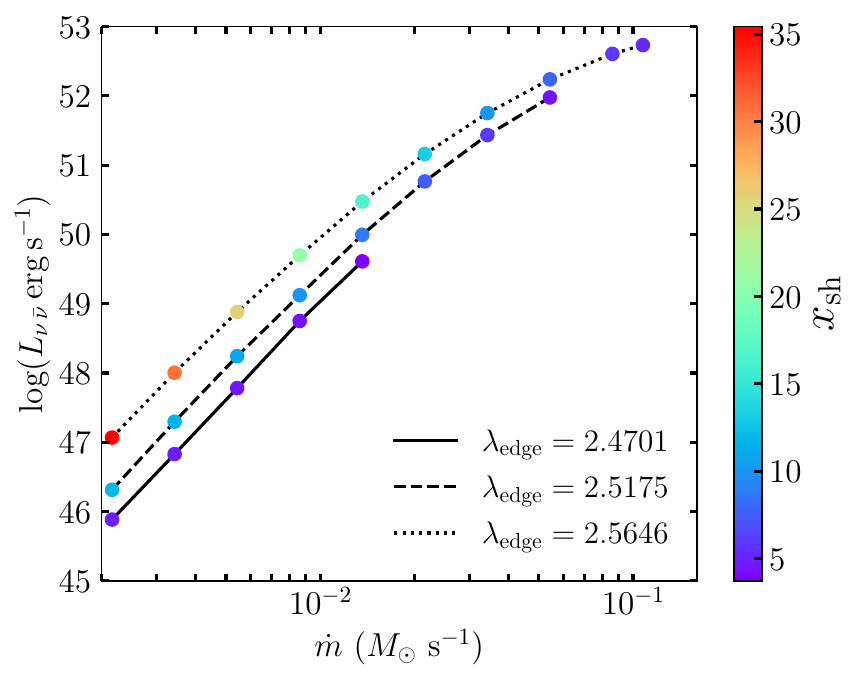}
    \caption{Variation of annihilation luminosity ($L_{\nu\bar{\nu}}$) with accretion rate ($\dot{m}$) for shocked solutions. Here, we fix the parameters as $(a_{\rm k }$, $\alpha$, $\varepsilon_{\rm edge}=0.99$, $0.001$, $4.834\times10^{-3})$. Filled circles connected by solid, dashed and dotted lines represent results corresponding to $\lambda_{\rm edge}=2.4701$, $2.5175$ and $2.5646$, respectively. Colors denote the shock radii. See the text for details.}
    \label{fig:fig05}
\end{figure}

Next, we compare the neutrino annihilation luminosity $L_{\nu\bar{\nu}}$ for flows injected with different $\lambda_{\rm edge}$ values. The obtained results are presented in Fig. \ref{fig:fig05}, where $L_{\nu\bar{\nu}}$ is plotted with $\dot m$. Here, we choose $\varepsilon_{\rm edge }=4.834\times 10^{-3}$, $\alpha =0.001$ and $a_{\rm k}=0.99$. In the figure, filled circles joined with solid, dashed and dotted lines denote results corresponding to $\lambda_{\rm edge}=2.4701$, $2.5175$ and $2.5646$, respectively. The colors indicate the shock radii with the range illustrated by the color bar. It is evident that $L_{\nu\bar{\nu}}$ increases with $\dot m$ for a fixed $\lambda_{\rm edge}$. Furthermore, when $\dot m$ is held constant, $L_{\nu\bar{\nu}}$ also increases for higher values of $\lambda_{\rm edge}$, as this enlarges the size of the post-shock region (PSC) due to the larger shock radius.

\begin{figure}
    \centering
       \includegraphics[width=\columnwidth]{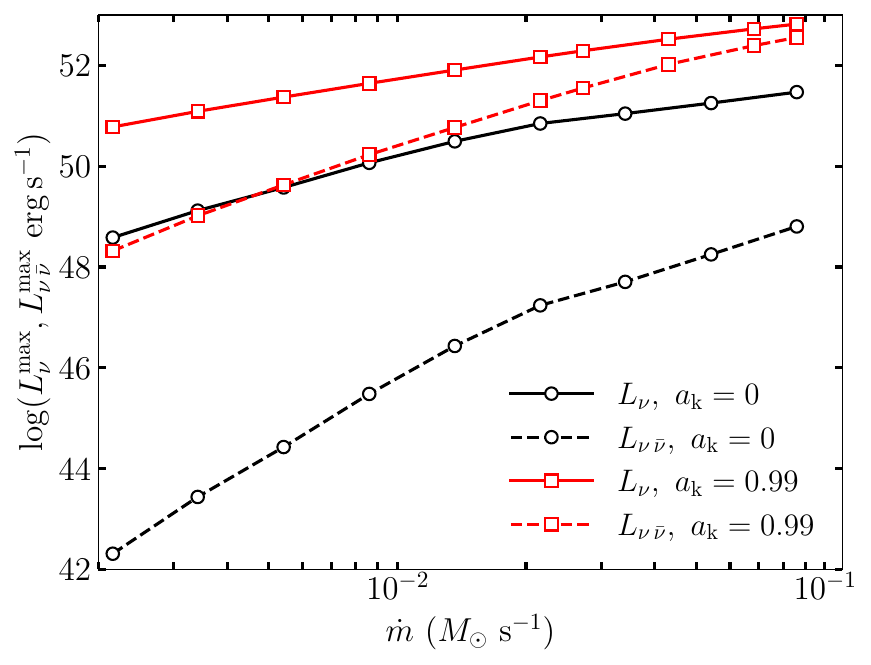}
    \caption{Variation of maximum neutrino luminosity ($L^{\rm max}_{\rm{\nu}}$) and neutrino annihilation luminosity ($L^{\rm max}_{\nu \bar{\nu}}$) with accretion rate $\dot{m}$ for different $a_{\rm k}$ values. Here, we choose viscosity parameter as $\alpha=0.005$. Open circles (black) connected by solid and dashed lines represent $L^{\rm max}_{\rm{\nu}}$ and $L^{\rm max}_{\nu \bar{\nu}}$ for a non-rotating ($a_{\rm k}=0$) black hole, whereas open squares (red) connected by solid and dashed lines indicate $L^{\rm max}_{\rm{\nu}}$ and $L^{\rm max}_{\nu \bar{\nu}}$ for a rotating ($a_{\rm k}=0.99$) black hole. See the text for details.}
    \label{fig:fig06}
\end{figure}

Thereafter, we put efforts to estimate the maximum neutrino luminosity $L_{\nu}^{\rm max}$ and maximum neutrino annihilation luminosity $L_{\nu \bar{\nu}}^{\rm max}$ using our model formalism. In order for that we choose $\alpha=0.005$, and compute $L_{\nu}^{\rm max}$ and $L_{\nu \bar{\nu}}^{\rm max}$ for a weakly rotating ($a_{\rm k}\rightarrow 0$) as well as rapidly rotating ($a_{\rm k}\rightarrow 1$) black holes by freely varying both $\varepsilon_{\rm edge}$ and $\lambda_{\rm edge}$. The obtained results are depicted in Fig. \ref{fig:fig06}, open circles and open squares joined with solid and dashed lines denote the variation of $L_{\nu}^{\rm max}$ and $L_{\nu \bar{\nu}}^{\rm max}$ with accretion rate $\dot m$ for $a_{\rm k}=0$ and $a_{\rm k}=0.99$, respectively. Overall, we observe that as the accretion rate increases, both luminosities rise. This is attributed to the enhancement of flow density, which allows matter to cool more efficiently. Additionally, we find that for a fixed $\dot m$, the luminosity is consistently higher for greater spin. From this analysis, we infer that the shock-induced global NDAF formalism is potentially promising for explaining the exceedingly high neutrino luminosity, as it is derived from a wide range of model parameters ($\alpha$, $\dot{m}$, $a_{\rm k}$).

\begin{figure}
    \centering
       \includegraphics[width=\columnwidth]{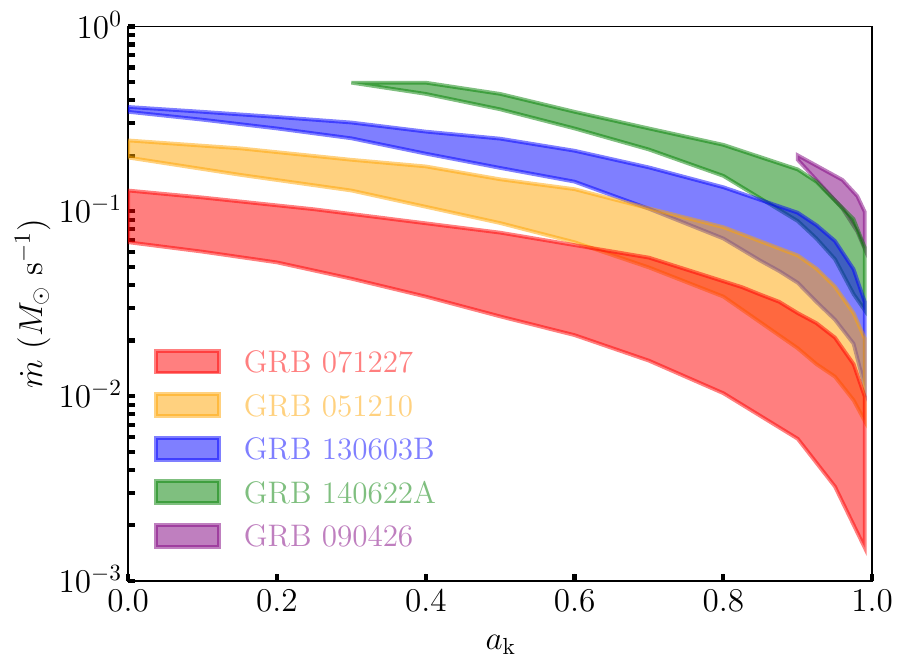}
    \caption{Effective domain of parameters in $a_{\rm k}-\dot{m}$ plane that satisfactorily explains $L^{\rm obs}_{\nu\bar{\nu}}$ for GRB sources 071227 (red), 051210 (orange), 130603B (blue), 140622A (green) and 090426 (purple) is illustrated. See the text for details. }
    \label{fig:fig07}
\end{figure}

\begin{table*}
    \caption{Observable parameters of GRBs under consideration. In columns $1-9$, GRB sources, burst duration ($T_{90}$), redshift ($z^\prime$), $E_{\rm \gamma,iso}$, $E_{\rm k,iso}$, opening angle ($\theta_{\rm j}$), GRB output power ($\dot{E}$), annihilation luminosity ($L^{\rm{obs}}_{\nu,\bar{\nu}}$) and references are presented. }
	\begin{ruledtabular}
 	      \begin{tabular}{lcccccccl}
            Source & $T_{90}$ & $z^\prime$  & $E_{\rm \gamma,iso}$ & $E_{\rm k,iso}$  & $\theta_{\rm j}$& $\dot{E}$  & $L^{\rm{obs}}_{\nu,\bar{\nu}}$& References\\
            & (s)      &    & ($\times 10^{51}$erg) & ($\times 10^{51}$ erg) &(rad) & (erg s$^{-1}$) & (erg s$^{-1}$)&  \\
 	      \hline \\
            GRB 071227  & 1.80     &  0.381    & 0.080    & 0.25   & $\sim 0.05$   & $3.1\times 10^{47}$ & $3.1\times 10^{48}$ & \cite{Liu-etal-2015}\\
            GRB 051210  & 1.30     &  1.300     & 0.360    & 2.38   & $\sim 0.05$   & $6.1\times 10^{48}$ & $6.1\times 10^{49}$& \cite{Liu-etal-2015}\\
            GRB 130603B & 0.18    &  0.356    & 0.200  & 2.80  & $\sim 0.07$ & $4.3\times 10^{49}$    & $4.3\times 10^{50}$& \cite{Fong-berger-2013}\\
            GRB 140622A  & 0.13     &  0.959    & 0.065    & 9.77   & $\sim 0.05$    & $3.3\times 10^{50}$ & $3.3\times 10^{51}$& \cite{Liu-etal-2015}\\
            GRB 090426 & 1.20     &  2.609   & 2.840    & 135.00   & $\sim 0.07$  & $1.9\times 10^{51}$   & $1.9\times 10^{52}$& \cite{Nicuesa-etal-2011}	
 		\end{tabular}	
	\end{ruledtabular}
 	\label{table}
 \end{table*}

\section{Astrophysical implications}
\label{astrophysical_implications}

In this section, we apply our model formalism to elucidate the energy output associated with the annihilation luminosity of gamma-ray bursts (GRBs). To achieve this, we select five GRBs for which the isotropic radiated energy during the prompt emission phase ($E_{\rm \gamma,iso}$), the isotropic kinetic energy of the outflow powering the afterglow phase ($E_{\rm k,iso}$), the opening angle of the ejecta $\theta_{\rm j}$, the duration of the burst ($T_{90}$), and the redshift ($z^\prime$) are known from the literature \cite[]{Nicuesa-etal-2011,Fong-berger-2013,Liu-etal-2015}. Using these observational data, we estimate the mean fireball output power from the central engine ($\dot{E}$) of the gamma-ray burst (GRB) as described in \cite[]{Fan-Wei-2011, Liu-etal-2015}, which is given by,
\begin{equation}
\dot{E} \approx \frac {(1+z^\prime)(E_{\rm \gamma,iso}+E_{\rm k,iso})\theta_{\rm j}^{2}}{2 T_{90}}.\label{dotE}
\end{equation}
For the selected gamma-ray bursts (GRBs) under consideration, we observe a wide range of energy release rates as $\dot{E}\sim 10^{47-52}$ erg s$^{-1}$ (see Table \ref{table}).

Meanwhile, the simulation results for GRBs suggets that $\dot{E}$ is a fraction of the neutrino annihilation luminosity. Consequently, the observed neutrino annihilation luminosity ($L^{\rm obs}_{\nu\bar{\nu}}$) of GRBs can be estimated as follows \cite[]{Aloy-etal-2005}:
\begin{equation}
L^{\rm obs}_{\nu\bar{\nu}} = \dot{E}/\eta,\label{mean_out_pow}
\end{equation}
where $\eta$ denote the efficiency of energy conversion by neutrino-antineutrino annihilation. It is worth noting that the plausible estimate of the $\eta$ value is not well-constrained due to persistent uncertainties in the efficiency \cite[]{Liu-etal-2015}. Therefore, we adopt a reasonable estimate of $\eta = 0.1$ for illustrative purposes. We present the observational properties of the GRB sources under consideration in Table \ref{table}.

To explain the observed neutrino annihilation luminosity $L^{\rm obs}_{\nu\bar{\nu}}$ of GRBs, we theoretically compute $L_{\nu\bar{\nu}}$ (see equation (\ref{neutrino_anni_lum})). Indeed, the precise estimates of mass ($M_{\rm BH}$), spin ($a_{\rm k}$), and accretion rate ($\dot m$) for GRB sources remain uncertain, as these quantities have not been conclusively determined. Hence, in this work, we adopt a reasonable estimate of source mass as $M_{\rm BH}=3M_\odot$, and compute $L_{\nu\bar{\nu}}$ across a range of $a_{\rm k}$ and $\dot m$. This is achieved by varying model parameters, including energy, angular momentum, and viscosity of the flow. The results obtained are illustrated in Fig. \ref{fig:fig07}, where the bounded regions in the $a_{\rm k}-{\dot m}$ plane correspond to the observed $L^{\rm obs}_{\nu\bar{\nu}}$ values for each GRB source. The region shaded in red, orange, blue, green, and purple are for GRB 071227 ($L^{\rm obs}_{\nu\bar{\nu}}=3.1 \times 10^{48}$ erg s$^{-1}$), GRB 051210 ($L^{\rm obs}_{\nu\bar{\nu}}=6.1 \times 10^{49}$ erg s$^{-1}$), GRB 130603B ($L^{\rm obs}_{\nu\bar{\nu}}=4.3 \times 10^{50}$ erg s$^{-1}$), GRB 140622A ($L^{\rm obs}_{\nu\bar{\nu}}=3.3 \times 10^{51}$ erg s$^{-1}$), and GRB 090426 ($L^{\rm obs}_{\nu\bar{\nu}}=1.9 \times 10^{52}$ erg s$^{-1}$), respectively. For GRB 071227, GRB 051210, and GRB 130603B, the computed range of accretion rates are $0.0015\lesssim\dot{m}\lesssim 0.13$, $0.0073 \lesssim\dot{m}\lesssim 0.24$ and $0.011 \lesssim \dot{m}\lesssim 0.37$, respectively, encompassing the spin range $0 \le a_{\rm k} < 1 $. In contrast, for GRB 140622A, the central source appears to be rotating either moderately or rapidly, with $0.3 \le a_{\rm k} < 0.99$ and an accretion rate in the range $0.03\lesssim\dot{m}\lesssim 0.5$. Furthermore, for GRB 090426, the source is found to be rapidly rotating, with $a_{\rm k} > 0.9$ and $0.06 \lesssim\dot{m}\lesssim 0.2$. These findings evidently suggest that rapidly rotating black holes seems to be more favorable in explaining the energy outputs of GRBs.

\section{\label{conclusions}Conclusions}

In this investigation, we explore steady, axisymmetric, neutrino-dominated accretion flows surrounding rotating black holes. By incorporating neutrino cooling ($Q_{\nu}$), we solve the governing equations in a self-consistent manner and identify transonic solutions, both with and without shocks, for a specified set of flow parameters ($\dot{m}, \alpha, a_{\rm k}, \varepsilon_{\rm edge}, \lambda_{\rm edge}$) that comply with the inner boundary condition. Additionally, we derive accretion solutions that manifest shock phenomena under the same parameter conditions. These shocked solutions seems to be more viable than the shock-free solutions as previously discussed in \cite{Becker-Kazanas-2001}. Our key findings are summarized below.

\begin{itemize}
    \item We find that global transonic neutrino-dominated accretion flows (NDAFs) exist that pass through either inner ($x_{\rm in}$) and outer ($x_{\rm out}$) critical points for the first time, to the best of our knowledge. Our analysis demonstrates that by varying the accretion rate $\dot{m}$ while keeping other flow parameters constant at the outer edge ($x_{\rm edge}$), the character of the solutions changes as they pass $x_{\rm out}$ rather than $x_{\rm in}$. Notably, we observe that close to the horizon, the temperature ($T$) and density ($\rho$) of the NDAF exceed $T \gtrsim 10^{10}$ K and $\rho \gtrsim 10^{8}$ g cm$^{-3}$ which are favorable for neutrino emission \cite[]{Popham-etal-1999} (see Fig. $\ref{fig:fig01}$).

    \item We obtain the shock-induced global transonic solutions in the realm of neutrino dominated accretion flows at remarkably high accretion rates (see Fig. \ref{fig:fig02}). Our findings indicate that the post-shock corona (PSC) is characterized by increased density and elevated temperatures compared to the pre-shock flow, creating a highly radiative environment conducive to substantial neutrino emission (see Fig. \ref{fig:fig03}).

    \item We also investigate the impact of $\dot{m},  \varepsilon_{\rm edge}, \lambda_{\rm edge}$ on the annihilation luminosity ($L_{\nu\bar{\nu}}$) for shocked solutions. Our analysis demonstrates that $L_{\nu\bar{\nu}}$ exhibits an increasing trend with $\dot{m}$ when flow parameters ($\alpha,  \varepsilon_{\rm edge}, \lambda_{\rm edge}, a_{\rm k }$) are held fixed (see Fig. \ref{fig:fig04} and Fig. \ref{fig:fig05}). This enhancement is attributed to the increased density of the flow as $\dot{m}$ increases, yielding more radiative environment in NDAFs.

    \item Furthermore, we compute the maximum neutrino luminosity $L_{\nu}^{\rm max}$ and maximum neutrino annihilation luminosity $L_{\nu \bar{\nu}}^{\rm max}$, examining how these quantities vary with accretion rate ($\dot{m}$) for weakly ($a_{\rm k} \rightarrow 0$) and rapidly ($a_{\rm k} \rightarrow 1$) rotating black holes (see Fig. \ref{fig:fig06}). Our findings reveal that both $L_{\nu}^{\rm max}$ and $L_{\nu\bar{\nu}}^{\rm max }$ exhibit a positive correlation with $\dot m$ for fixed values of $a_{\rm k}$, and that higher $a_{\rm k}$ results in enhanced luminosities for a given accretion rate.

    \item We employ our model formalism to explain the neutrino annihilation luminosity $L^{\rm obs}_{\nu\bar{\nu}}$ (see equation (\ref{mean_out_pow})) observed from the gamma-ray bursts (GRBs). We find that shock-induced global neutrino-dominated accretion flow (NDAF) model model satisfactorily accounts $L_{\nu\bar{\nu}}$, corresponding to the energy release rates of five GRBs, which range from $\dot{E}\sim 10^{47-52}$ erg s$^{-1}$. Based on these findings, we infer the potential range of accretion rates for the GRBs, considering parameters, such as  $a_{\rm k}$ and $M_{\rm BH}$. We find that for $0 \le a_{\rm k} < 1 $ and $M_{\rm BH}=3M_\odot$, the accretion rate for GRB 071227, GRB 051210, and GRB 130603B are $0.0015\lesssim\dot{m}\lesssim 0.13$, $0.0073 \lesssim\dot{m}\lesssim 0.24$ and $0.011 \lesssim \dot{m}\lesssim 0.37$, respectively. For GRB 140622A, we obtain $0.03\lesssim\dot{m}\lesssim 0.5$ for $0.3 \le a_{\rm k} < 0.99$ and $M_{\rm BH}=3M_\odot$. On contrary, the central source of GRB 090426 ($M_{\rm BH}=3M_\odot$) appears to be rapidly rotating ($a_{\rm k} > 0.9$) and accreting within the range $0.06 \lesssim\dot{m}\lesssim 0.2$ (see Fig. \ref{fig:fig07}). 
    
\end{itemize}

Finally, we mention that for the first time to the best of our knowledge, we explore the  shock-induced global NDAFs around rotating stellar mass black holes to explain the neutrino annihilation luminosity of gamma-ray bursts (GRBs). It is important to note that the present formalism relies on several approximations. We adopt an effective potential to mimic the spacetime geometry around a rotating black hole, rather than working within a general relativistic framework. We do not consider magnetic fields, even though they are ubiquitous in all astrophysical environments. We ignore the effects of self-gravity as well. Implementation of these physical processes is beyond the scope of this study and we intend to investigate them in future research.

\section*{Data Availability}
 
The data underlying this article will be available with reasonable request.

\section*{Acknowledgments}

Work of AK is support by Council and Scientific \& Industrial Researc, India. SC thanks MATRICS, Science and Engineering Research Board (SERB), India, for support through grant MTR/2022/000318. SD acknowledges financial support from MATRICS, Science and Engineering Research Board (SERB), India, through grant MTR/2020/000331. 

\bibliography{reference}

\appendix

\section{Expressions for $\mathcal{N}, \mathcal{D}, \lambda_{11},\lambda_{12},\Theta_{11},\Theta_{12}$}

The equations of radial momentum (\ref{rad_mom_eq}), mass conservation (\ref{mass_cons_eq}), azimuthal momentum (\ref{azimuthal_eq}) and entropy generation equation (\ref{entrop_gen_eq}) can be combinedly expressed as,
\begin{equation}
\mathcal{A}_{\lambda}\frac{d\lambda}{dx} +\mathcal{A}_{\Theta}\frac{d\Theta}{dx} +\mathcal{A}_{v}\frac{dv}{dx}+\mathcal{A}_{x}=0, \label{LE_CL}
\end{equation}
\begin{equation}
\mathcal{B}_{\lambda}\frac{d\lambda}{dx} +\mathcal{B}_{\Theta}\frac{d\Theta}{dx} +\mathcal{B}_{v}\frac{dv}{dx}+\mathcal{B}_{x}=0, \label{LE_CS}
\end{equation}
\begin{equation}
\mathcal{C}_{\lambda}\frac{d\lambda}{dx} +\mathcal{C}_{\Theta}\frac{d\Theta}{dx} +\mathcal{C}_{v}\frac{dv}{dx}+\mathcal{C}_{x}=0, \label{LE_CE}
\end{equation}
where the coefficients $\mathcal{A}_{j},\mathcal{B}_{j}$ and $\mathcal{C}_{j}$, $j \rightarrow \lambda, \Theta, v, x$, in the above equations are functions of $x$, $v$, $\lambda$, and $\Theta$. Using equations (\ref{LE_CL}), (\ref{LE_CS}) and (\ref{LE_CE}), we obtain the equations (\ref{dvdr}), (\ref{dldr}) and (\ref{dthetadr}) and the coefficients involved in these equations are given by,
\begin{align*}
    & \qquad \qquad \qquad \mathcal{N}=\mathcal{C}_{x}\mathcal{A}_{\Theta}\mathcal{B}_{\lambda} - \mathcal{C}_{\Theta}\mathcal{A}_{x}\mathcal{B}_{\lambda} - \mathcal{C}_{x}\mathcal{A}_{\lambda}\mathcal{B}_{\Theta} + \mathcal{C}_{\lambda}\mathcal{A}_{x}\mathcal{B}_{\Theta} + \mathcal{C}_{\Theta}\mathcal{A}_{\lambda}\mathcal{B}_{x} - \mathcal{C}_{\lambda}\mathcal{A}_{\Theta}\mathcal{B}_{x},\\
    & \qquad \qquad \qquad \mathcal{D}=
    -\mathcal{C}_{v}\mathcal{A}_{\Theta}\mathcal{B}_{\lambda} + \mathcal{C}_{\Theta}\mathcal{A}_{v}\mathcal{B}_{\lambda} + \mathcal{C}_{v}\mathcal{A}_{\lambda}\mathcal{B}_{\Theta} - \mathcal{C}_{\lambda}\mathcal{A}_{v}\mathcal{B}_{\Theta} - \mathcal{C}_{\Theta}\mathcal{A}_{\lambda}\mathcal{B}_{v} + \mathcal{C}_{\lambda}\mathcal{A}_{\Theta}\mathcal{B}_{v}, \\
    &\lambda_{11}=\frac{\mathcal{A}_{x}\mathcal{B}_{\Theta} - \mathcal{A}_{\Theta}\mathcal{B}_{x}}{\mathcal{A}_{\Theta}\mathcal{B}_{\lambda} - \mathcal{A}_{\lambda}\mathcal{B}_{\Theta}},\quad   \lambda_{12}=\frac{\mathcal{A}_{v}\mathcal{B}_{\Theta} - \mathcal{A}_{\Theta}\mathcal{B}_{v}}{\mathcal{A}_{\Theta}\mathcal{B}_{\lambda} - \mathcal{A}_{\lambda}\mathcal{B}_{\Theta}},\quad   \Theta_{11}=\frac{-\mathcal{A}_{x}\mathcal{B}_{\lambda} + \mathcal{A}_{\lambda}\mathcal{B}_{x}}{\mathcal{A}_{\Theta}\mathcal{B}_{\lambda} - \mathcal{A}_{\lambda}\mathcal{B}_{\Theta}},\quad  \Theta_{12}=\frac{-\mathcal{A}_{v}\mathcal{B}_{\lambda} + \mathcal{A}_{\lambda}\mathcal{B}_{v}}{\mathcal{A}_{\Theta}\mathcal{B}_{\lambda} - \mathcal{A}_{\lambda}\mathcal{B}_{\Theta}}.
\end{align*}
where
\begin{align*}
    & \qquad \qquad \qquad \mathcal{A}_\lambda =1 + \lambda_{1} \Theta_{1}, \quad   \mathcal{A}_\Theta =\lambda_{1} \Theta_{2}, \quad \mathcal{A}_{v}= \frac{A_1S_1}{v~C_{s0}}, \quad\mathcal{A}_{x}=\lambda_{2}+\lambda_{1}\Theta_{3},\\
    & \qquad \qquad \qquad \mathcal{B}_{\lambda}= R_{1} + R_{2} \Theta_{1},\quad \mathcal{B}_{\Theta}= R_{2} \Theta_{2},\quad \mathcal{B}_{v}= R_{3} + R_{2} \Theta_{4},\quad \mathcal{B}_{x}= R_{4} + R_{2} \Theta_{3},\\
    & \qquad \qquad \qquad \mathcal{C}_{\lambda} =  \mathcal{T}_1 -C_{s}^2  v \mathcal{T}_2, \quad \mathcal{C}_{\Theta} =  \mathcal{T}_3 -C_{s}^2  v \mathcal{T}_4, \quad \mathcal{C}_{v} =  \mathcal{T}_5 -C_{s}^2  v \mathcal{T}_6,  \quad  \mathcal{C}_{x}= \mathcal{T}_7 -C_{s}^2  v \mathcal{T}_8, \\
    & \lambda_{1} =-\frac{2 C_{s} x\alpha}{v}, \quad \Theta_{1}=\frac{-C_s A_1 S_1\mathcal{F}_1 }{2 \mathcal{F}(2 C_{s}^2- A_1 S_1)}, \quad A_1 = \frac{11 \pi\bar{a} c^3 M_{BH}^2 m_{\rm e}^2 }{3 k_{\rm B} \dot{M}} , \quad S_1= v H\Theta^4 \sqrt{\Delta}, \quad \mathcal{F}_{1}= g_1 \frac{(\lambda\ \Omega_\lambda+\Omega)}{(1-\lambda \Omega)^2},\\
    & g_1=\frac{(x^2+a^2_{\rm{k}})^2+2 \Delta a^2_{\rm{k}}}{(x^2+a^2_{\rm{k}})^2-2 \Delta a^2_{\rm{k}}}, \quad 
    \Omega_\lambda =\frac{\partial \Omega}{\partial \lambda}, \quad C_{\rm s}= \frac{Y_{1}+ \sqrt{Y_{1}^2 + 4 Y_{2}} }{2}, \quad Y_{1}= A_1 v  \sqrt{\frac{ \Delta x^3}{\mathcal{F}}} \Theta^4, \quad Y_{2}= \frac{m_{\rm e}}{m_{\rm p }}\Theta, \\
    & \lambda_{2}=\frac{\alpha(C_{\rm s}^2+v^2)}{v}\left(\frac{x}{2}\frac{\Delta^{\prime}}{ \Delta}-2\right), \quad \Delta^{\prime}=\frac{d\Delta}{dx},  \quad \Theta_{3}=\frac{A_1 S_1}{ C_{s0}}\left(\frac{3}{2 x} - \frac{\mathcal{F}_{2}}{2\mathcal{F}} \right) + \frac{S_1 A_1 \Delta^{\prime}}{ 2 C_{s0} \sqrt{\Delta} }, \quad C_{s0}= 2 C_s -\frac{A_1 S_1}{C_s}, \\
    & \mathcal{F}_{2}= \frac{1}{1-\lambda\Omega} \left(\frac{d g_1}{d x} + \gamma_{\phi}^2 \lambda g_1 \frac{d\Omega}{dx}\right) , \quad R_{1}= C_{\rm s}^2 \frac{\mathcal{F}_{1}}{2\mathcal{F}}, \quad R_{2}= C_{\rm s},\quad R_3= v-\frac{C_{\rm s}^2}{v}, \quad \Theta_{4}= \frac{A_1 S_1 }{v C_{s0}},\\
    & R_{4}= \frac{d \Phi^{ \rm{eff}}}{dx} +C^2_{\rm s}\left(\frac{\mathcal{F}_2}{2 \mathcal{F}}- \frac{\Delta^{\prime}}{2 \Delta} -\frac{3}{2x}\right), \quad \mathcal{T}_1= \Gamma_1  -3A_1S_1 v \left( \frac{\mathcal{F}_1}{2 \mathcal{F}}  +\frac{\Theta_{1}}{C_s} \right), \quad \Gamma_1= -\alpha(v^2 + C_{s}^2)x\Omega_\lambda ,\\
    & \mathcal{T}_2= \frac{\mathcal{F}_{1}}{2\mathcal{F}} -\frac{\Theta_{1} }{C_s} , \quad \mathcal{T}_3  =  \frac{ v}{\gamma-1} \frac{m_{\rm e}}{m_{\rm p}} + 3A_1 S_1 v \left(\frac{4}{\Theta} +\frac{\Theta_2}{C_s}\right), \quad  \mathcal{T}_4=  -\frac{\Theta_{2}}{C_{s}}, \quad \mathcal{T}_5 =  3 A_1 S_1   +\frac{3 A_1 S_1 \Theta_{4} v}{C_s}, \quad \mathcal{T}_6 =  -\frac{\Theta_4}{C_s} -\frac{1}{v},\\
    & \mathcal{T}_7 =  \Gamma_2- \frac{Q_\nu}{\Sigma} + \mathcal{C}_{1},~\Gamma_2= -\alpha(v^2 + C_{s}^2)x \frac{d\Omega}{dx}, \quad \mathcal{C}_{1}= \frac{3 A_2 S_1 v}{2} \left(\frac{\Delta^{\prime}}{\Delta}- \mathcal{F} \mathcal{F}_2 + \frac{2 \Theta_3}{C_s} +\frac{3}{x} \right),\quad \mathcal{T}_8 = -\frac{\Theta_3}{C_s} +\frac{\mathcal{F}_2}{2 \mathcal{F}}- \frac{\Delta^{\prime}}{2 \Delta} -\frac{3}{2x}.
\end{align*}

\section{Constants in energy deposition rate $l_{\nu\bar{\nu}}$}

In equation (\ref{lnunubar}), the constants $A_{1,i}$ and $A_{2,i}$ are given by \cite[]{Ruffert-etal-1997,Rosswog-etal-2003},
\begin{align*}
      & A_{{1,i}}= \frac{\sigma_{0}}{ 12 \pi^2 c (m_{\rm{e}} c^2)^2} \left[ \left( C_{V,\nu_{i}} - C_{{A,\nu_{i}}} \right)^2 + \left( C_{{V,\nu_{i}}} + C_{{A,\nu_{i}}} \right)^2 \right] \quad {\rm and} \quad A_{{2,i}}=\frac{\sigma_{0}}{6 \pi^2 c} \left( 2 C_{{V,\nu_{i}}}^2 - C_{{A,\nu_{i}}}^2 \right),
\end{align*}
where 
$\sigma_{0}=1.76\times 10^{-44}~\rm cm^2$, $\quad C_{V,\nu_{\rm e}}=\frac{1}{2}+2\sin^2 \theta_{W}$, $\quad C_{V,\nu_{\mu}}=C_{V,\nu_{\tau}}=-\frac{1}{2}+2 \sin^2 \theta_{ W}$, $\quad C_{A,\nu_{\rm e}}=C_{ A,\bar{\nu}_{\mu}}=C_{{A},\bar{\nu}_{\tau}}=\frac{1}{2}$, $\quad C_{A,\bar{\nu}_{\rm e}}=C_{ A,\nu_{\mu}}=C_{\rm{A},\nu_\tau}=-\frac{1}{2} \quad$ and $\quad\sin^2 \theta_{ W}=0.23$.

\section{Method of calculating accretion solution with shock}
 
Here, we describe the methodology employed to obtain the shock solution.

\begin{enumerate}

\item For an accretion solution originating at $x_{\rm edge}$ and passing through the outer critical point ($x_{\rm out}$), we evaluate the total pressure $\Pi$, accretion rate $\dot M$, and local energy $\varepsilon$ using the supersonic flow variables at a radial coordinate $x$, where $x < x_{\rm out}$.

\item We make use of RHCs ($i.e.$ the conservation of $\Pi$, $\dot M$ and $\varepsilon$) to obtain three algebraic equations and solve them to calculate subsonic flow variables.

\item Using these subsonic flow variables, we numerically integrate equations (\ref{dvdr}-\ref{dthetadr}) towards the black hole and check the critical point conditions (equation \ref{N=D=0}).

\item If critical point conditions are satisfied, we obtain the inner critical point ($x_{\rm in}$, see Fig. \ref{fig:fig02}) and use the inner critical point flow variables to integrate equations (\ref{dvdr}-\ref{dthetadr}) all the way to the horizon $x_{\rm h}$. This yields a complete global accretion solution with a shock connecting $x_{\rm h}$ and $x_{\rm edge}$.

\item When critical point conditions are not satisfied, we decrease the radial coordinate $x$ and repeat the process (point 1-4) until a complete accretion solution with a shock is found.\\

\end{enumerate}

\end{document}